\documentclass[aps, prb, twocolumn, superscriptaddress]{revtex4}

\usepackage{graphicx}
\usepackage{ifpdf}
\ifpdf
\usepackage[
  bookmarks=true,
  bookmarksopen=true,
  breaklinks=true,
  colorlinks=true,
  linkcolor=blue,anchorcolor=blue,
  citecolor=blue,filecolor=blue,
  menucolor=blue,pagecolor=blue,
  urlcolor=blue]{hyperref}
\else
\usepackage[
  breaklinks=true,
  colorlinks=true,
  linkcolor=blue,anchorcolor=blue,
  citecolor=blue,filecolor=blue,
  menucolor=blue,pagecolor=blue,
  urlcolor=blue]{hyperref}
\fi

\begin{document}

\title{Dissipative solitons stabilized by a quantum Zeno-like effect}
\author{Hong-Gang Luo}
\affiliation{Center for Interdisciplinary Studies, Lanzhou
University, Lanzhou 730000, China} \affiliation{Key Laboratory for
Magnetism and Magnetic Materials of the Ministry of Education,
Lanzhou University, Lanzhou 730000, China} \affiliation{Institute
of Theoretical Physics, Chinese Academy of Sciences, Beijing
100080, China}
\author{Dun Zhao}
\affiliation{School of Mathematics and Statistics, Lanzhou
University, Lanzhou 730000, China} \affiliation{Center for
Interdisciplinary Studies, Lanzhou University, Lanzhou 730000,
China}
\author{Xu-Gang He}
\affiliation{School of Mathematics and Statistics, Lanzhou
University, Lanzhou 730000, China}
\author{Lin Li}
\affiliation{Center for Interdisciplinary Studies, Lanzhou
University, Lanzhou 730000, China}


\maketitle

{\sffamily The solitary wave\cite{Russell1834} (or the
soliton\cite{ZK1965}) is a ubiquitous nonlinear phenomenon, which
has been widely observed in many systems, particularly, in optical
fibers \cite{Hasegawa1973, Mollenauer1980} and Bose-Einstein
condensates (BEC) \cite{Burger1999, Strecker2002, Becker2008}. It
was recognized that the stability of a soliton is due to a subtle
balance between competing features like dispersion and
nonlinearity in a medium. The practical applications of the
soliton raise the requirement for optimal soliton management
\cite{Taylor1992, Serkin2000, Malomed2006}. However, due to the
presence of dissipation, the soliton inevitably fades away. Here
we show that the dissipative effect can be completely suppressed
by an analogue of the quantum Zeno effect \cite{Misra1977} (in
short we will call it quantum Zeno-like effect) and propose a
novel way to stabilize the soliton. The proposal opens a new
perspective to realize an ideal optical soliton transmission, a
seemingly impossible dream in communication technologies. It is
also of fundamental interest to the quantum Zeno effect itself and
the matter-wave dynamics.}

In quantum mechanics, the quantum Zeno effect shows that an
unstable particle can be completely stabilized by frequent
measurements \cite{Misra1977}, which has been confirmed by many
experiments \cite{Itano1990, Fischer2001, Streed2006} and its
possible applications have been extensively explored
\cite{Hosten2006, Erez2008}. Solitons are localized, finite energy
states in a medium. They behave like elementary particles, can
pass through each other and preserve their shapes and speeds after
collisions. However, the presence of dissipation effect makes the
soliton behave like an unstable particle, so it will inevitably
decay and eventually disappears in a medium. The insight gained
from the quantum Zeno effect in quantum mechanics provides a
possible mechanism to stabilize such a soliton if it is frequently
``measured". Through a rigorous mathematical analysis, the soliton
is, indeed, shown to become stable if the external potential
applied and/or the Feshbach resonance along with the dispersion is
periodically modulated with high frequency. This is due to the
fact that the ``inevitable" dissipation effect is completely
suppressed by such a frequent ``measurement", even in the presence
of the dissipation in reality.

We study the dynamics of the soliton governed by a one-dimensional
(1D) dimensionless nonautonomous nonlinear Schr\"odinger (NLS)
equation
\begin{widetext}
\begin{equation}
i\frac{\partial u(x,t)}{\partial t} + \varepsilon\,
f(x,t)\frac{\partial ^2 u(x,t)}{\partial x^2} + \delta\,
g(x,t)|u(x,t)|^2 u(x,t) - \frac{1}{2} V(t) x^2 u(x,t) =
i\frac{\gamma(t)}{2} u(x, t). \label{non-eq1}
\end{equation}
\end{widetext}
Here $u(x,t)$ can represent either the envelope of pulses in
nonlinear optics with different notations or the macroscopic wave
function of BEC with strong transverse confinement (i.e.,
effective 1D model) at the mean-field level. Therefore, the
soliton satisfying equation (\ref{non-eq1}) can be the temporal
(if V(t) = 0) or the spatial optical soliton or the matter-wave
soliton. For convenience, below we use the notations for the
matter-wave solitons. $f(x,t)$ and $g(x,t)$ are dimensionless and
denote the time- and space-dependent dispersion and nonlinearity
managements, respectively. Here the space- and time-coordinates
$x$ and $t$ are measured in units of $a_r$ and $\omega_r^{-1}$,
where $a_r$ and $\omega_r$ being the transverse harmonic
oscillator length and the transverse confining frequency of BEC,
respectively. $\varepsilon$ and $\delta$ are constants in the
standard NLS equation (see Method). $V(t)$ represents the external
harmonic potential applied, whose strength and sign can be tuned
experimentally. The linear dissipation ($\gamma < 0$) or gain
($\gamma > 0$) rate is also time-dependent. Generally, these
coefficients should be real. Due to the introduction of these
control parameters, the soliton obeying equation (\ref{non-eq1})
can be called as nonautonomous \cite{Serkin2007} or dissipative
\cite{Akhmediev2005} soliton or similariton in nonlinear optics
\cite{Kruglov2000, Dudley2007, Ponomarenko2007}.

The integrability study of the nonautonomous NlS equation has a
long history, see, for example, ref. (\onlinecite{Serkin2000}). A
complete integrability condition of equation (\ref{non-eq1})
neglecting dissipation and/or gain (i.e., $\gamma(t) \equiv 0$)
was clearly shown in ref.(\onlinecite{Serkin2007}) and further
studied recently by the Painlev\'e analysis \cite{Weiss1983,
Ablowitz1991, Zhao2008}. However, it is nontrivial to explore the
integrability of equation (\ref{non-eq1}) in the presence of the
dissipation and/or gain effect. Below we show a generalized
``integrability" condition of equation (\ref{non-eq1}). It can be
obtained by the Painlev\'e analysis \cite{Weiss1983}. For details
of the Painlev\'e analysis one can refer to
ref.(\onlinecite{Ablowitz1991}). It reads
\begin{widetext}
\begin{eqnarray}
&& -2\varepsilon f(t) V(t) = \frac{1}{f(t)}\frac{d^2}{dt^2}f(t) -
\frac{1}{f^2(t)}\left(\frac{d}{dt}f(t)\right)^2 -
\frac{1}{g(t)}\frac{d^2}{dt^2}g(t) +
\frac{2}{g^2(t)}\left(\frac{d}{dt}g(t)\right)^2 \nonumber \\
&& \hspace{2.5cm} -\frac{1}{f(t)}\left(\frac{d}{dt}f(t)\right)
\frac{1}{g(t)}\frac{d}{dt}g(t) -
\left(\frac{1}{f(t)}\frac{d}{dt}f(t) -
\frac{2}{g(t)}\frac{d}{dt}g(t)\right) \gamma(t) -
\frac{d}{dt}\gamma(t) + \gamma^2(t). \label{integrability}
\end{eqnarray}
\end{widetext}
Here we would like to point out that the dispersion and
nonlinearity managements are not allowed to be space-dependent,
even if their space-dependence assumptions are initially made in
equation (\ref{non-eq1}). This is a necessary condition for
equation (\ref{non-eq1}) to pass through the Painlev\'e test.
Hereafter the coordinate $x$ in $f(x, t)$ and $g(x, t)$ is
omitted. In addition, we also mention that equation
(\ref{integrability}) is essentially not a complete integrability
condition in a strict physical sense, namely, the infinite
conservation laws are not assumed. It is the condition under which
the Painlev\'e analysis can pass through. In this sense, we will
call it the Painlev\'e integrability condition, which sheds a new
light on the ``integrability" behavior of equation
(\ref{non-eq1}).

The seemingly complicated Painlev\'e integrability condition
obtained for the first time shows a subtle balance between four
parameters $f(t), g(t), V(t)$ and $\gamma(t)$ affecting the
dynamics of the nonautonomous soliton, and thus has a profound
implication to the control of the soliton dynamics. Any three
parameters out of them can be set independently, while the
remaining one can be tuned according to equation
(\ref{integrability}). In this case, the soliton satisfied
equation (\ref{non-eq1}) can be managed as a whole in the sense of
the Painlev\'e integrability.  This feature provides many
possibilities to control the soliton dynamics in different fields
such as nonlinear optics and BEC, deserving further exploration.

Here we focus on the dynamical stability of the soliton, which is
of fundamental importance to the study of the matter-wave dynamics
and in particular to the propagation of the optical solitons in
transmission lines. In this case, $f(t), g(t)$ and/or $V(t)$ are
set independently and $\gamma(t)$ is effectively determined by
these three independent parameters. At first glance, this
assumption seems to be logically incorrect since the dissipation
and/or gain mechanism always physically exists in reality.
However, when some control parameters are modulated, the {\it
actual} $\gamma(t)$ acting on solitons might be different from
that without such modulations. Here it is quite instructive to
think of a comparison with the quantum Zeno effect. A physically
unstable particle always decays with a natural decay rate named as
$R_0$. However, it is well known that that frequent measurements,
for example, with a frequency $\omega$, on the unstable particle
can slow down its decay. This fact tells us that the actual
dissipation rate named as $R_\omega$ of the unstable particle with
frequent measurements must be different from $R_0$. In this case,
$R_0$ is only a nominal, while $R_\omega$ is the actual
dissipation rate of the unstable particle. The soliton case is
similar. The only difference is that the soliton is not a real
particle, though it behaves like an elementary particle. For
convenience, we take the dissipation and/or gain as
$\gamma_\omega(t)$ when dispersion, nonlinearity or the external
potential is modulated periodically with frequency $\omega$.

Focusing on the dynamical stability of the soliton, we rewrite
equation (\ref{integrability}) as
\begin{equation}
\frac{d}{dt}\gamma_{\omega}(t) - \gamma^2_{\omega}(t) -
\alpha(t)\gamma_{\omega}(t) + \beta(t) = 0, \label{eq-gamma}
\end{equation}
where $\alpha(t) = \frac{1}{f(t)}\frac{d}{dt}f(t) -
\frac{2}{g(t)}\frac{d}{dt}g(t)$ and $\beta(t)$ represents the
remaining terms in (\ref{integrability}) independent of
$\gamma_{\omega}(t)$. This is a standard Riccati equation
\cite{Boyce1986}. Mathematically, the Riccati equation can not be
solved analytically in a general case. Here we resort to the
Runge-Kutta method to solve numerically this equation. We first
consider the effect of the external potential $V(t)$ by fixing
$f(t)$ and $g(t)$ to $f(t) = g(t) = \pm 1$ and let $\varepsilon =
1/2$ as usual. Thus, $\alpha(t) = 0$ and $\beta(t) = \mp V(t)$.

\begin{figure}[tbp]
\includegraphics[width=\columnwidth, angle=0]{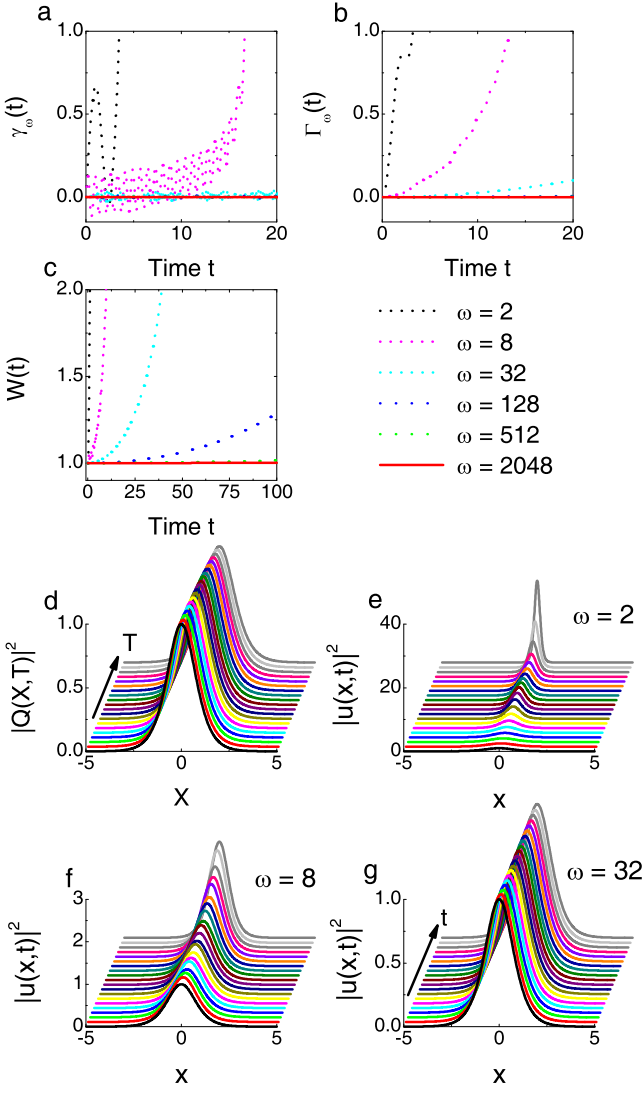}
\caption{{\bf The effective dissipation and/or gain rates and the
time evolution of the nonautonomous solitons in the periodically
modulated external potentials.} {\bf a}, The effective dissipation
and/or gain rate $\gamma_{\omega}(t)$. {\bf b}, The effective
accumulated dissipation and/or gain rate, which is defined by
$\Gamma_{\omega}(t) = \int_0^t \gamma_{\omega}(t')dt'$. {\bf c},
The normalized overlap integral defined by $W(t) = \frac{\int
u^*(x, 0) u(x, t) dx}{\int |u(x, 0)|^2 dx}$. Here $u(x, t)$ obeys
equation (\ref{non-eq1}) and $u(x,0)$ is its initial state, which
is equivalent to the canonical bright soliton. {\bf d}, The
canonical bright soliton $Q(X,T) = \mbox{sech}(X)\exp(iT/2)$. {\bf
e - f}, The explicit evolution of the nonautonomous bright soliton
in periodically modulated harmonic external potentials with
increasing frequencies from $\omega = 2$ to $32$. The time
interval for the soliton evolution is from 0 to 10 (4 in the case
of $\omega = 2$). Here the frequency and time are measured in
units of $\omega_r$ and $\omega_r^{-1}$, respectively. The
strength of the harmonic external potential is modulated by $V(t)
= V\cos(\omega t)$ with $V = 1$. Other parameters used are $f(t) =
g(t) = 1$, $\varepsilon = 1/2$ and for all cases the initial value
of $\gamma_{\omega}(t)$ is set to be zero. The result of $f(t) =
g(t) = -1$ is similar. } \label{fig1}
\end{figure}

\begin{figure}[h]
\includegraphics[width=\columnwidth, angle=0]{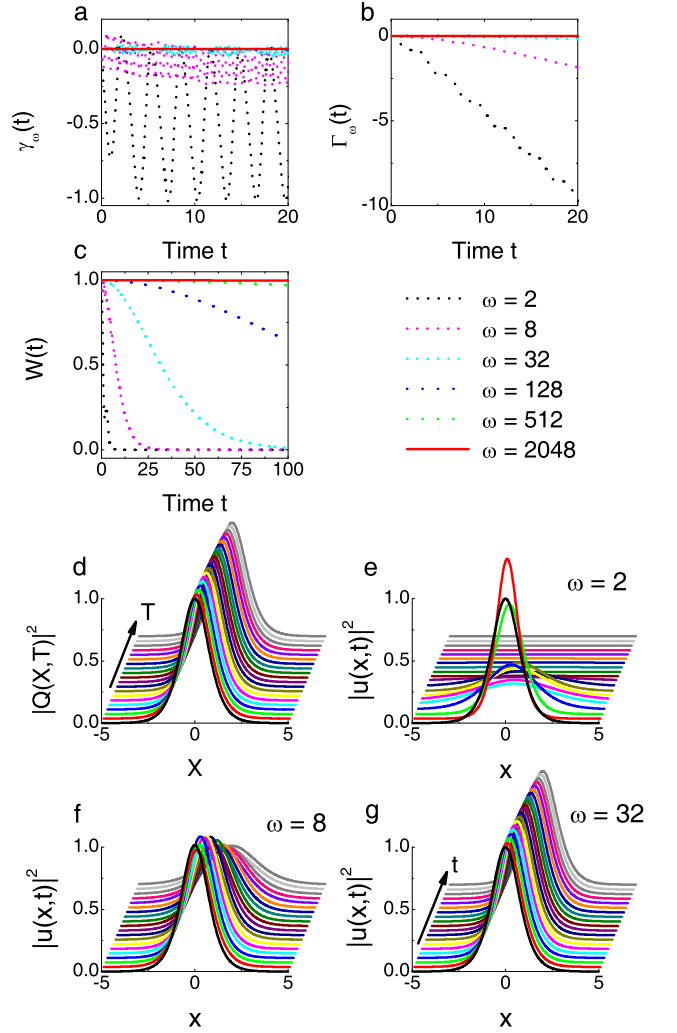}
\caption{{\bf The effective dissipation or gain rates and the time
evolution of the nonautonomous solitons with periodically
modulated nonlinearity.} {\bf a - g} fully correspond to those in
Figure \ref{fig1}. The time interval for the soliton evolution in
{\bf e-g} is plotted from 0 to 10. The nonlinearity modulation is
given by $g(t) = \exp[\delta a \cos(\omega t)]$, where $\delta a =
1/\omega^2$ is taken. Other physical variables used are $V(t) = 0$
and $f(t) = 1$. } \label{fig2}
\end{figure}

Figure \ref{fig1}{\bf a} shows $\gamma_{\omega}(t)$ as a function
of time. At low frequencies, $\gamma_{\omega}(t)$ oscillates and
turns slowly up with time. Finally, at certain moment
$\gamma_{\omega}(t)$ diverges. With increasing frequency, the
diverging point of $\gamma_{\omega}(t)$ is postponed rapidly and
finally, at the time interval we studied (up to $100$ here and not
fully shown for clarity) no divergence is observed at least for
$\omega > 128$. Meanwhile, the magnitude of oscillation decreases
quickly. This clearly indicates that the matter-wave soliton can
be stabilized by applying a modulated external potential with high
frequency. This can also be seen from the behavior of
$\Gamma_{\omega}(t)$ with increasing frequencies, as shown in
Figure \ref{fig1}{\bf b}. When $\omega = 2048$, no visible
deviation from zero is observed for the whole time domain studied.
The explicit time evolution of the nonautonomous bright soliton is
shown in Figure \ref{fig1}{\bf e-f} with frequencies of $\omega =
2, 8$, and $32$. With increasing frequencies, the nonautonomous
bright soliton approaches gradually to the canonical bright
soliton shown in Figure \ref{fig1}{\bf d}. The overlap integral of
the evolving bright soliton with its initial state is shown in
Figure \ref{fig1}{\bf c} for different frequencies. At high
frequencies the soliton evolves slowly and the deviation from its
initial state keeps invisible. Analytically, the nonautonomous
soliton can be effectively reduced to the canonical soliton form
since $\Gamma_{\omega}\rightarrow 0$ at the limit of high
frequency. As a result, $a(x,t) = 0, X(x,t) = x, T(t) = t,
\mbox{and}\, c(t) = 0$ neglecting the constants $C_1, C_2$, and
$C_3$ (see Method). In this case, the Painlev\'e integrability
condition recovers consistently to the complete integrability
condition \cite{Serkin2007, Zhao2008}. As a result, the
nonautonomous soliton becomes stable and completely integrable.
This is a quantum Zeno-like effect in the context of nonlinear
dynamics found here for the first time. It builds up a clear
relationship between the nonautonomous and the canonical solitons.
This surprising result unambiguously shows the particle nature of
the nonautonomous solitons and provides a dynamical way to
stabilize them.

Experimentally, it is easy to modulate the external potential by
changing periodically the direction and magnitude of the magnetic
field controlling the magnetic traps for BEC. Actually, an easier
way to stabilize the matter-wave solitons is to use the Feshbach
resonance management\cite{Kevrekidis2003}. To show this, let $f(t)
= 1$ and $g(t) = \exp(\delta a \cos(\omega t))$, where $\delta a$
and $\omega$ are the amplitude and the frequency of the Feshbach
resonance modulation. For simplicity, we also take $V(t) = 0$. In
this case, one has
\begin{eqnarray}
&& \alpha(t) = -2 \delta a\, \omega \sin(\omega t), \nonumber \\
&& \beta(t) = - \delta a\, \omega^2 \left(\cos(\omega t) - \delta
a \sin^2(\omega t)\right). \nonumber
\end{eqnarray}
Limiting to weak disturbances, we take $\delta a = 1/\omega^2$.
The result is shown in Figure \ref{fig2}{\bf a \& b}. Similarly,
the matter-wave soliton can be stabilized by a high frequency
modulation Feshbach resonance. As an example, Figure \ref{fig2}
{\bf e-f} shows the nonautonomous bright solitons evolving with
time. Likewise, with increasing frequencies, the nonautonomous
soliton approaches gradually to the canonical soliton evolution,
as shown in Figure \ref{fig2}{\bf d}. For $\omega = 2048$, the
nonautonomous soliton has no visible deviation from its initial
state, as shown in Figure \ref{fig2}{\bf c}. This is another
example that the nonautonomous soliton can be stabilized by a
quantum Zeno-like effect. Theoretically, it is also possible to
modulate periodically the dispersion to stabilize the
nonautonomous soliton, which is quite readily realized in the
context of optical soliton transmission.

The above discussions are entirely based on the Painlev\'e
integrability condition (\ref{integrability}) and are not limited
to the bright soliton solution. In fact, any allowed solution of
the standard NLS equation, including the bright and dark solitons,
the multi-soliton solutions, and even the periodic plane wave
solutions, can be stabilized by such a scheme. This provides a
novel approach to study in detail the matter-wave properties and
related soliton dynamics.

The same scheme can also be applied to the optical soliton. The
present result has an important implication that with the help of
the quantum Zeno-like effect a dissipationless, ideal optical
soliton transmission might be possible in reality. Once it is
realized, the impact to modern information technology is indeed
far-reaching.

The above discussion is based on an intuitive physics
consideration and the Painlev\'e integrability condition obtained
by a rigorous mathematical derivation. One may question its
realizability, even be suspicious of it as an artifact. In fact,
some previous results have an implication that our result is
physically reasonable. For example, the earlier numerical
simulations of the nonautonomous NLS equation in BEC
\cite{Saito2003, Abdullaev2003} have clearly indicated that the 2D
bright soliton can be stabilized for some time by tuning
periodically the nonlinearity, even if the instability of
multidimensional solitons is eventually inevitable
\cite{Konotop2005}. A recent experiment in optics
\cite{Centurion2006} also clearly demonstrated that the
propagation of femtosecond pulses was stabilized in the layered
Kerr media consisting of glass and air (the nonlinearity
management). Loosely speaking, the fact that the dispersion
management in nonlinear optics can effectively improve the optical
soliton transmission quality is also compatible to the present
result. It is expectable that an appropriately designed dispersion
management can further improve the optical soliton transmission
quality, and ideally yield a dissipationless transmission.

\vspace{0.5cm} \noindent {\bf Method}

Once the Painlev\'e integrability condition equation
(\ref{integrability}) is satisfied, an exact analytical solution
of the nonautonomous NLS equation (\ref{non-eq1}) can be obtained
from the corresponding canonical solution of the standard NLS
equation
\begin{equation}
i\frac{\partial }{\partial T}Q(X,T)+\varepsilon \frac{\partial
^2}{\partial X^2}Q(X,T)+\delta \left\vert Q(X,T)\right\vert
^{2}Q(X,T) = 0. \label{nls}
\end{equation}
Here $\varepsilon$ and $\delta$ are constants. When $\delta
\varepsilon > 0$ ($< 0$), the standard NLS equation (\ref{nls})
has bright (dark) soliton solutions. This is materialized by a
general transformation \cite{Zhao2008}
\begin{equation}
u(x,t) = Q(X(x,t),T(t))e^{ia(x,t) + c(t)}, \label{trans1}
\end{equation}
where $X(x,t), T(t), a(x,t)$ and $c(t)$ are real functions to be
determined by the requirement that $u(x,t)$ and $Q(X,T)$ satisfy
equations (\ref{non-eq1}) and (\ref{nls}), respectively. Inserting
equation (\ref{trans1}) into equation (\ref{non-eq1}) and
comparing with equation (\ref{nls}), we obtain a set of ordinary
differential equations for these transformation parameters, whose
solutions read
\begin{eqnarray}
&& a(x,t) = \frac{1}{4\varepsilon f(t)}\left(\frac{d}{dt}\ln\left(
\frac{f(t)}{g(t)}\right) - \gamma(t)\right)\,x^2 \nonumber
\\
&& \hspace{0.2cm}+ C_1 \frac{g(t)}{f(t)}e^{\Gamma(t)}\,x  - C_1^2
\varepsilon\, \int d\,t'\,
\frac {g(t')^2}{f(t')} e^{2\Gamma(t')} + C_2,\label{trans2}\\
&& X(x,t)= \frac{g(t)}{f(t)}e^{\Gamma(t)} x - 2\varepsilon C_1
\int dt' \,
\frac{g(t')^2}{f(t')} e^{2\Gamma(t')}. \label{trans3} \\
&& T(t) = \int dt' \frac{g^2(t')}{
f(t')} e^{2\Gamma(t')} + C_3, \label{trans4} \\
&& c(t) = \frac{1}{2}\ln \frac{g(t)}{f(t)} + \Gamma(t),
\label{trans5}
\end{eqnarray}
where $\Gamma(t) = \int_0^t\gamma(t')dt'$ is accumulated
dissipation and/or gain effect. $C_1, C_2$, and $C_3$ are
arbitrary constants and are set to be zero for simplicity. Through
this transformation, all exact solutions, including the canonical
solitons, of the standard NLS equation can be recast into the
corresponding solutions of the nonautonomous NLS equation
(\ref{non-eq1}). This result yields a mapping between the
canonical and the nonautonomous solitons in a straightforward way.

{\bf Acknowledgement} We are grateful to L. Yu, C.-P. Sun, M.-L.
Du, W.-M. Liu, W.-M. Zheng for helpful discussions and V. V.
Konotop for valuable correspondence. Support from the NSFC, the
national program for basic research, and the program for NCET of
China is acknowledged.


\end{document}